\documentclass[conference]{IEEEtran}
\IEEEoverridecommandlockouts
\usepackage{hyperref}
\usepackage{float}
\usepackage{cite}
\usepackage{subcaption}
\usepackage{multirow}
\usepackage{amsmath,amssymb,amsfonts}
\usepackage{algorithmic}
\usepackage{graphicx}
\usepackage{textcomp}
\usepackage{soul}
\usepackage{xcolor}
\usepackage{comment}
\usepackage{array}
\usepackage[compatibility=false]{caption}
\usepackage{tabularx}

\makeatother

\begin{document}

\title{From the Two-Capacitor Paradox to Electromagnetic Side-Channel Mitigation in Digital Circuits\\
}

    
\author{\IEEEauthorblockN{Raghvendra Pratap Singh$^1$, Baibhab Chatterjee$^2$, Shreyas Sen$^3$, Debayan Das$^1$}
\IEEEauthorblockA{\textit{$^1$Dept. of Electronic Systems Engineering, Indian Institute of Science, Bangalore, India} \\}
\IEEEauthorblockA{\textit{$^2$Dept. of ECE, University of Florida, Gainesville, USA} \\}
\IEEEauthorblockA{\textit{$^3$Dept. of ECE, Purdue University, USA} \\}
\thanks{This work was supported in part by the Department of Science \& Technology (DST), India (DST/ENCASE/P-1/2024), NSF SaTC Award
CNS-2343128, and the Cyber Security Karnataka (CySecK) initiative by the Govt of Karnataka, India. } 
}

\maketitle
\begin{abstract}
The classical two-capacitor paradox of the lost energy is revisited from an electronic circuit security standpoint. The paradox has been solved previously by various researchers, and the energy lost during the charging of capacitors has been primarily attributed to the heat and radiation. We analytically prove this for various standard resistor-capacitor (RC) and resistor-inductor-capacitor (RLC) circuit models. From the perspective of electronic system security, electromagnetic (EM) side-channel analysis (SCA) has recently gained significant prominence with the growth of resource-constrained, internet-connected devices. This article connects the energy lost due to capacitor charging to the EM SCA leakage in electronic devices, leading to the recovery of the secret encryption key embedded within the device. Finally, with an understanding of how lost energy relates to EM radiation, we propose adiabatic charging as a solution to minimize EM leakage, thereby paving the way towards low-overhead EM SCA resilience.

\end{abstract}


\vspace{-2mm}
\section{Introduction}\label{sec:intro}
Electromagnetic side-channel analysis (EM SCA) exploits unintentional radiation from integrated circuits (ICs) to extract sensitive information such as cryptographic keys \cite{gandolfi2001electromagnetic}. This leakage originates from the switching of the transistors and the resulting displacement currents through the metal-interconnect stack in a 3-D IC, which act as antennas and radiate \cite{das2020and, nath2021multipole, das2019stellar}.

\begin{figure}[!t]
\centerline{\includegraphics[scale=0.65]{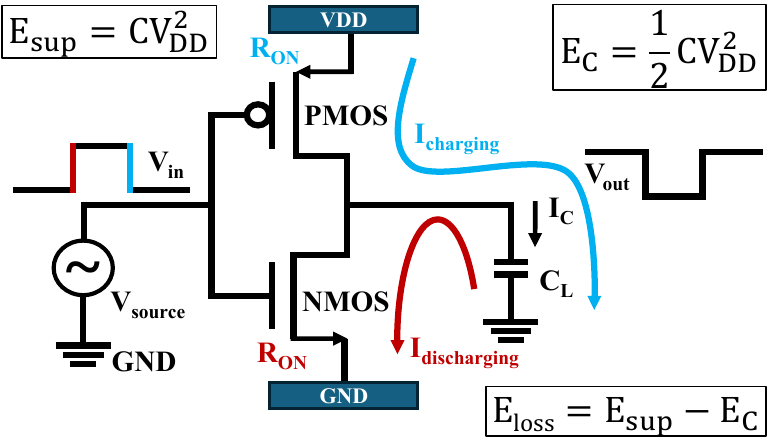}}
\vspace{-1mm}
\caption{Transistor circuits like CMOS inverter operation during signal transitions. When the input switches, either the PMOS or the NMOS conducts, causing the load capacitor \( C \) to charge or discharge. The resulting transient current leads to energy dissipation through heat and EM radiation.}
\vspace{-3mm}
\label{inverter}
\end{figure}

To understand the physical source of this leakage, consider a complementary metal-oxide semiconductor (CMOS) inverter charging a load capacitor \( C \) to a logic high voltage \( V_{DD} \) (see Fig.~\ref{inverter}). The energy stored in the capacitor after full charging is \(E_C = \frac{1}{2} C V_{DD}^2\). As depicted in Fig.~\ref{inverter}, when an input signal transitions in a digital circuit, the corresponding transistors switch states to alternately charge or discharge the load capacitor. The PMOS network conducts during charging, while the NMOS path enables discharging, with each path introducing real resistances that include both conductive and radiative components, such as the radiative resistance $R_{\text{rad}}$ \cite{balanis2016antenna}. The transient current flow in digital operations can be accurately modeled using RC or RLC circuits \cite{powell1979two, choy2004capacitors}. Each switching event generates both electric and magnetic fields, leading to heat loss and the radiation of EM energy \cite{kocher1999differential}. 

Fundamental RC/RLC circuit modeling reveals that the total dissipative energy per transition is always $E_{\text{heat}} + E_{\text{rad}} = \frac{1}{2} C V_{DD}^2$, distributed between heat and EM radiation \cite{boykin2002two}. This is discussed in detail in Section~\ref{section3}. Through a detailed analysis of the energy lost during capacitor charging, this paper establishes a clear link between this lost energy and the EM SCA attacks on ICs.   

\vspace{-1mm}
\section{Related Work }\label{section2}
\vspace{-1mm}
EM SCA attacks effectively recover secret keys from real-world embedded devices by correlating measured EM leakage during encryption with a leakage model \cite{lisovets_lets_2021, lomne_side_2021}. To address such attacks, the research community has proposed a wide range of solutions, from layout-level mitigations to algorithmic masking \cite{das2019stellar, nath2021multipole, das2022sca}. 

\begin{figure}[!t]
\centerline{\includegraphics[scale=0.5]{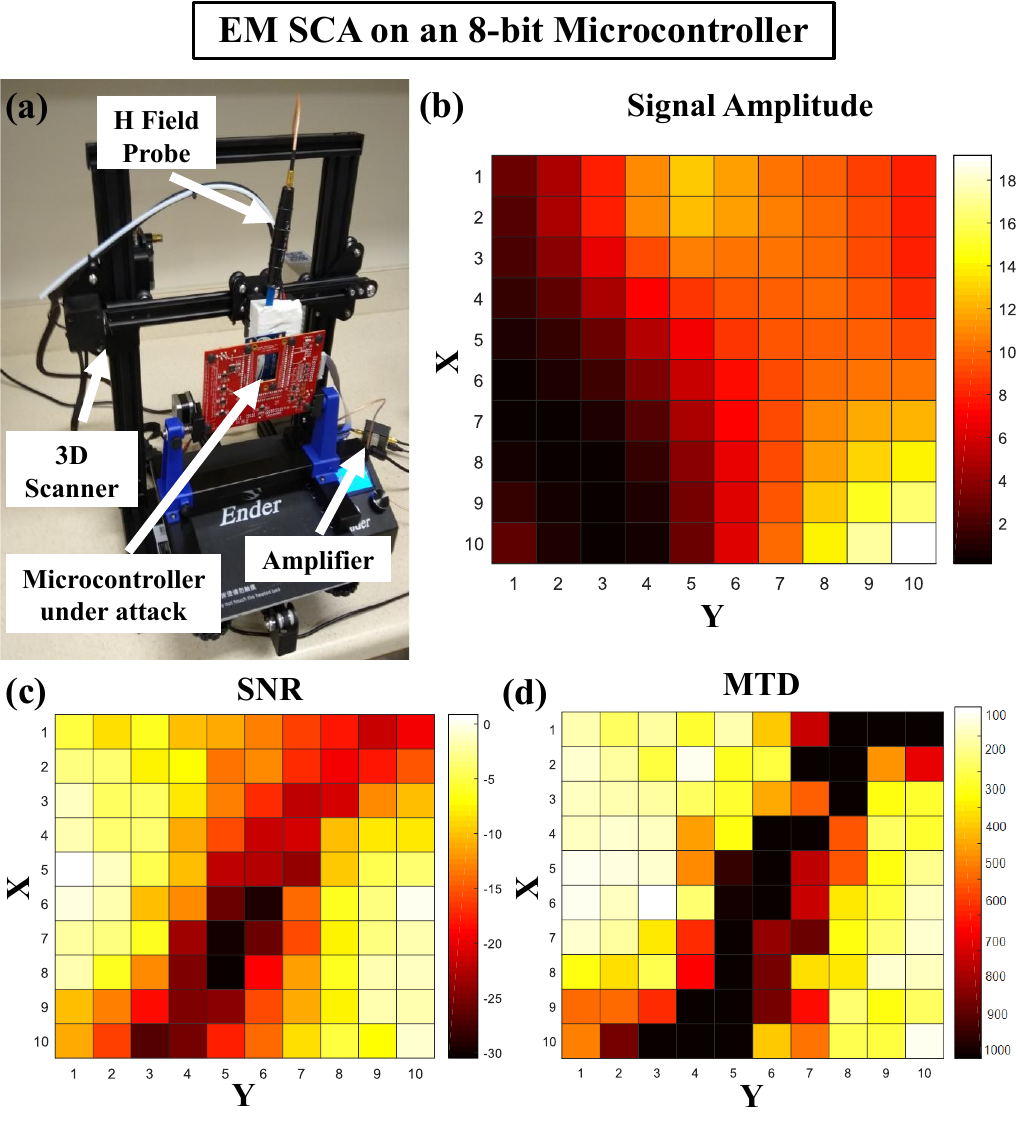}}
\vspace{-2mm}
\caption{EM SCA on an 8-bit Microcontroller: (a) Automated 3D scanning system with H-field probe and amplifier, (b) Heatmap of signal amplitude, (c) SNR map highlighting regions of strong EM leakage, and (d) MTD indicating regions with the highest data-dependent leakage \cite{danial2020scniffer}.}
\vspace{-6mm}
\label{scniffer}
\end{figure}

Fig.~\ref{scniffer}(a) depicts the EM SCA attack set-up to recover the secret key from a target 8-bit Atmel microcontroller running the AES-128 encryption algorithm. As the device performs encryption, traces are collected using the H-field probe mounted on a 3D scanner from different locations across the target microcontroller chip. Fig.~\ref{scniffer}(b) shows the signal amplitude heatmap ($10 \times 10$) across the surface of the target device under attack. Fig.~\ref{scniffer}(c) depicts the side-channel signal-to-noise ratio (SNR) heatmap scanning across the $10 \times 10$ grids. We can clearly see that the signal amplitude does not directly correlate with the SNR, as the microcontroller also performs operations other than encryption. Fig.~\ref{scniffer}(d) shows the minimum traces to disclosure (MTD) or the minimum number of encryption traces to recover the secret key. Note that the MTD heatmap is correlated to the side-channel SNR, and a lower MTD reveals an easier attack at that location.


Various logic and circuit-level countermeasures have been proposed to prevent EM SCA attacks. Among the circuit-level countermeasures, STELLAR \cite{das2019stellar} introduced a root-cause-driven methodology that combines local low-level metal routing with a signature attenuation hardware \cite{das_asni_2018} to mitigate side-channel leakage. Authors in \cite{das2022sca} proposed a reduced-leakage standard cell design using white-box pre-silicon analysis. Multipole routing techniques have been developed to spatially arrange current elements within the metal layers, thereby enhancing physical-layer security against EM SCA attacks \cite{nath2021multipole}.

In this work, we analyze the lost energy during the charging cycle of a transistor. From the fundamental physics viewpoint, we ask the following questions: Where does the energy go during digital switching?  What is the distribution of that energy between heat and radiation emitted from an IC? Lastly, how is this related to EM SCA attacks?

The well-known two-capacitor paradox demonstrates that when a charged capacitor (with energy \(\tfrac{1}{2} C V^2\)) is connected to an identical, uncharged capacitor, the post-equilibrium total stored energy drops to \(\tfrac{1}{4} C V^2\) \cite{powell1979two}. This reduction in stored energy has led researchers to investigate where the missing energy goes. While classical explanations \cite{powell1979two} identified this loss as Joule heating, more complete models have shown that the energy loss is partitioned between resistive and radiative losses \cite{boykin2002two, choy2004capacitors, singal2013paradox}. This paper connects the missing energy in switching circuits in the form of EM radiation to the EM SCA attacks. It examines how energy loss in transistor circuits, modeled as RC and RLC circuits, contributes to radiative leakage and how this understanding can help mitigate the EM SCA attacks on secure digital ICs.  

\section{Analysis of the Lost Energy During Capacitor Charging}\label{section3}
\vspace{-1mm}
\begin{figure*}[!t]
    \centering
    \includegraphics[width=\textwidth]{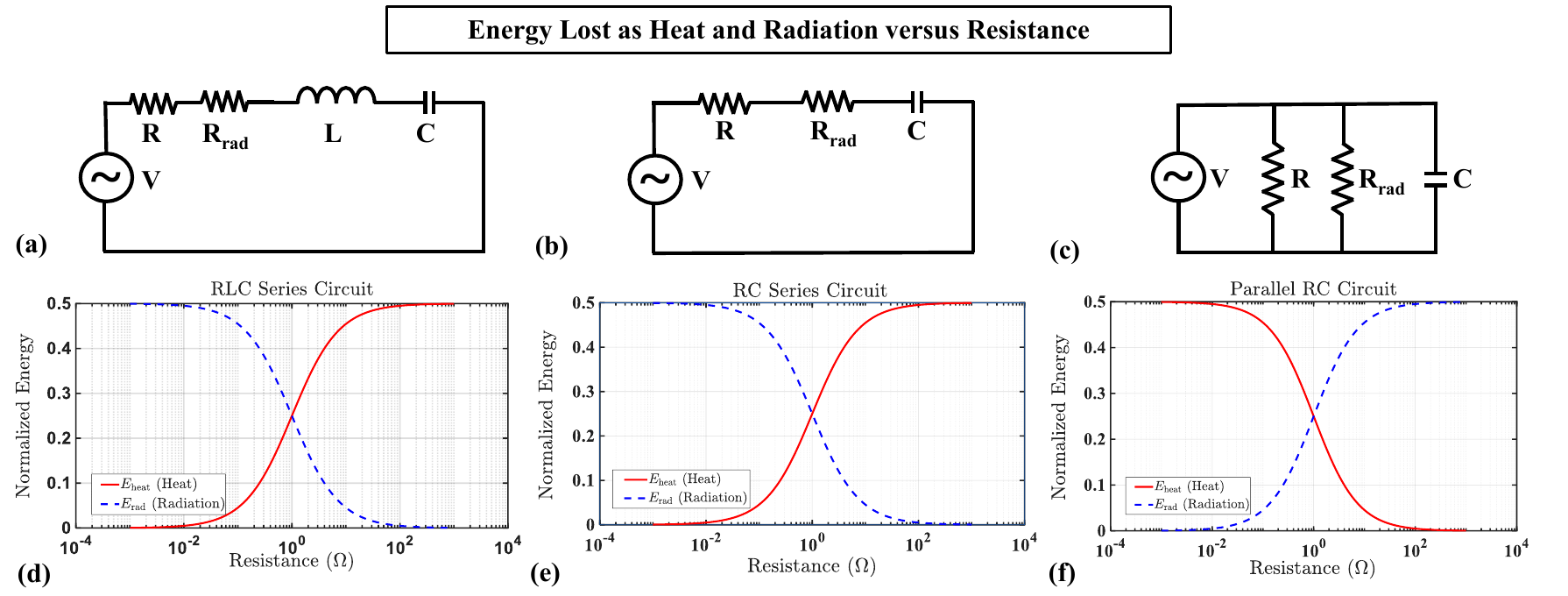}
    \vspace{-5mm}
    \caption{Energy lost as heat (\(E_{\text{heat}}\), red) and radiation (\(E_{\text{rad}}\), blue) versus resistance \(R\). Left to right: (a) Series RLC circuit, (b) Series RC circuit, and (c) Parallel RC circuit; respective energy plots are shown in (d), (e), and (f) plots. The figure illustrates how energy dissipation transitions from radiative to resistive loss as \(R\) varies across different circuit configurations.}
    \vspace{-5mm}
    \label{fig:energy_partition}
\end{figure*} 

In digital circuits, logic gates switch between high and low voltage levels. During a switching event, transistors turn on or off, and a capacitive load \(C\) at the output node either charges or discharges. This activity is the primary source of energy loss through heat and radiation in CMOS systems.

To model this behavior, we approximate the circuit as an ideal series RC or RLC network. These models help analyze how initial electrical energy is redistributed into stored charge, heat, and EM radiation.

\subsection{Heat Dissipation in the Standard RC Model}
\vspace{-1mm}
Assume a capacitor is charged from 0 to \(V_{DD}\) through a resistor \(R\). During the charging, the total energy supplied by the source is \( E_{\text{sup}} = \int_0^{t} V_{DD} \cdot i(t) \, dt = \int_0^{V_{DD}} {V_{DD}} \cdot C \, dV = C V_{DD}^2 \), while the energy stored in the capacitor is \( E_{\text{C}} = \int_0^{V_{DD}} V \cdot C \, dV = \frac{1}{2} C V_{DD}^2 \). Thus, half the energy is dissipated as heat or radiation. This finding is well described in the two-capacitor paradox problem, in which half of the energy appears to vanish when two capacitors share charge. Back in 1979, Powell \cite{powell1979two} demonstrated that in an RC circuit, half of the energy drawn inevitably dissipates as heat.

\subsection{EM Radiation During Capacitor Charging} \label{3b}
\vspace{-1mm}
While traditional RC models dissipate energy to resistive (Joule) losses, further studies have shown that a portion of this energy can also be radiated away as EM waves \cite{boykin2002two, choy2004capacitors}. In the case of practical circuits, especially those involving fast switching, inductive and radiative effects become non-negligible. These systems are modeled using RC and RLC circuits by including (\(R_{\text{rad}}\)) to account for EM wave emission.

To model radiative losses, a radiation resistance can be included in series with the conducting path. For small interconnects or capacitor structures (short dipole approximation), the radiation resistance is expressed as \(R_{\text{rad}} \approx 80\pi^2 \left(\frac{\ell}{\lambda}\right)^2 \) \cite{balanis2016antenna}, where \(\ell\) is the effective current path length and \(\lambda\) is the wavelength corresponding to the switching frequency.

Modeling the circuit with both resistive and radiative components leads to the governing differential equation:
\begin{equation}
    \frac{dV}{dt} + \frac{1}{(R + R_{\mathrm{rad}})C} V = 0,
\end{equation}
with the initial condition \(V(0) = V_0\), where \(R\) is the ohmic resistance and \(R_{\mathrm{rad}}\) is the effective radiation resistance representing power lost as EM waves. The solution to this differential equation is \(V(t) = V_0 e^{-t/\tau}\), where \(\tau = ({R + R_{\mathrm{rad}}})C\). The resulting current in the circuit is \(I(t) = -C ({dV}/{dt}) = ({C V_0}/{\tau}) e^{-t/\tau}\). The instantaneous power dissipated as heat and radiated power are \(P_{\mathrm{heat}}(t) = R I^2(t)\) and \(P_{\mathrm{rad}}(t) = R_{\mathrm{rad}} I^2(t)\), respectively. Integrating over time yields the total energies lost to heat and radiation:
\begin{equation}
    E_{\mathrm{heat}} = R \int_0^{\infty} I^2(t) \, dt, \quad
    E_{\mathrm{rad}} = R_{\mathrm{rad}} \int_0^{\infty} I^2(t) \, dt,
\end{equation}
and thus their sum is
\begin{equation}
    E_{\mathrm{heat}} + E_{\mathrm{rad}} = (R + R_{\mathrm{rad}}) \int_0^{\infty} I^2(t) \, dt.
\end{equation}

Calculating the integral,
\begin{equation}
    \int_0^{\infty} I^2(t) \, dt = \left(\frac{C V_0}{\tau}\right)^2 \int_0^{\infty} e^{-2t/\tau} dt = \left(\frac{C V_0}{\tau}\right)^2 \cdot \frac{\tau}{2} = \frac{C^2 V_0^2}{2 \tau},
\end{equation}
and substituting this back gives
\begin{equation}
    E_{\mathrm{heat}} + E_{\mathrm{rad}} = (R + R_{\mathrm{rad}}) \frac{C^2 V_0^2}{2 \tau}.
\end{equation}
Using the expression for \(\tau\), it follows that
\begin{equation}
    E_{\mathrm{heat}} + E_{\mathrm{rad}} = (R + R_{\mathrm{rad}}) \frac{C^2 V_0^2}{2({R + R_{\mathrm{rad}}})C} = \frac{1}{2} C V_0^2,
\end{equation}
which matches the initial energy stored on the capacitor \cite{boykin2002two, choy2004capacitors}. This derivation illustrates that the total dissipated energy during capacitor charge sharing is always equal to \((1/2)C V_0^2\), distributed between Joule heating and EM radiation losses.

\subsection{Energy Loss Distribution Across Different Circuit Models}
\vspace{-1mm}
In the previous subsection, we showed that the lost energy is distributed between heat and radiation. Fig. \ref{fig:energy_partition} illustrates how these energy components vary across different circuit topologies as a function of the load or series resistance \(R\). The resistive losses correspond to Joule heating, while the radiation losses model energy carried away by EM waves, represented mathematically by introducing an equivalent radiation resistance \(R_{\mathrm{rad}}\) \cite{boykin2002two, choy2004capacitors}. When \(E_0 = \frac{1}{2} C V_{DD}^2\) denotes the initial energy, the following formulas describe the normalized energy dissipation in various models:

\begin{itemize}
    \item \textbf{RLC Series Circuit:} 
    Fig. \ref{fig:energy_partition}(a) shows the RLC series circuit, where the inductor and capacitor set the natural frequency $\omega_0 = 1/\sqrt{LC}$ and the damping factor $\zeta = \tfrac{(R+R_{\mathrm{rad}})}{2}\sqrt{C/L}$. These parameters influence whether the transient response is underdamped or overdamped but do not change the fact that all of the initial energy $E_0 = \tfrac{1}{2} C V_0^2$ is eventually dissipated in $R$ and $R_{\mathrm{rad}}$. Using the RLC\,+\,$R_{\mathrm{rad}}$ framework of~\cite{boykin2002two}, the \(E_{\text{heat}}\) and \(E_{\text{rad}}\) curves plotted against resistance in Fig.~\ref{fig:energy_partition}(d) are derived by numerically solving the transient response for different $R$ and integrating $R i^2(t)$ and $R_{\mathrm{rad}} i^2(t)$ over time. These results confirm that as $R \rightarrow 0$ the loss is predominantly radiative, while for $R \gg \sqrt{L/C}$ it is dominated by Joule heating.

    \item \textbf{RC Series Circuit:}
    Here, energy loss occurs due to the interplay between the physical \(R\) and the \(R_{\mathrm{rad}}\), up to 50\% of $E_0$ is lost as heat for high values of $R$, with radiation loss dominating at very low $R$. Fig. \ref{fig:energy_partition}(b) showed the RC series circuit and Fig. \ref{fig:energy_partition}(e) illustrated the plot between \(E_{\text{heat}}\) and \(E_{\text{rad}}\) versus resistance. In this case, the normalized heat and radiation energies are given by:
    \begin{equation}
        E_{\mathrm{heat}} = E_0 \cdot \frac{R}{R + R_{\mathrm{rad}}}, \quad
        E_{\mathrm{rad}} = E_0 \cdot \frac{R_{\mathrm{rad}}}{R + R_{\mathrm{rad}}}
    \end{equation}

    \item \textbf{Parallel RC Circuit:}  
    In parallel configurations, the circuit shown in Fig. \ref{fig:energy_partition}(c) and Fig. \ref{fig:energy_partition}(f) illustrated the plot between \(E_{\text{heat}}\) and \(E_{\text{rad}}\) versus resistance. 

    The capacitor discharges through two parallel branches, a conductive path $R$ and a radiative path $R_{\mathrm{rad}}$, giving an effective resistance $R_p = \left(\frac{1}{R} + \frac{1}{R_{\mathrm{rad}}}\right)^{-1}$. Solving the transient and integrating the branch powers yields
    \begin{equation}
        E_{\mathrm{heat}} = E_0 \cdot \frac{R_{\mathrm{rad}}}{R + R_{\mathrm{rad}}}, \quad
        E_{\mathrm{rad}} = E_0 \cdot \frac{R}{R + R_{\mathrm{rad}}}
    \end{equation}
    In the parallel RC case, the heat loss remains below $0.5E_0$ for all $R$, illustrating the sharing of total losses between the concurrent conductive and radiative current paths.
\end{itemize}

These relationships emphasize that the total energy lost during charging remains consistent and equal to the initial capacitor energy \(E_0\), but its partitioning varies with circuit topology and resistance. Understanding this partitioning is essential, as the \(E_{\mathrm{rad}}\) represents a direct physical source of EM side-channel leakage exploitable in hardware security attacks. 

\section{Minimizing the EM Radiation for EM SCA Attack Resilience}
In conventional CMOS charging, a fast transition from 0 to \( V_{DD} \) induces large displacement currents and transient magnetic fields, leading to EM radiation. These fast transitions are especially vulnerable to EM SCA, as the emitted radiation correlates strongly with internal data and switching activity \cite{das2019stellar}. This section discusses techniques to mitigate such emissions, focusing on adiabatic charging \cite{athas1994low, dickinson2002adiabatic}.



In adiabatic charging, assume a linear voltage ramp from 0 to \( V_{DD} \) over time \( T \). The resulting current is constant, \(I(t) =  C \cdot \frac{V_{DD}}{T}\). As charging time \(T\) increases, both the resistive power dissipation and the EM radiation (which depends on \( dI/dt \)) decrease inversely with \(T\) \cite{dickinson2002adiabatic, athas1994low}:

\begin{equation}
E_{\text{heat}} = \frac{C^2 V_{DD}^2 R}{T}; \quad E_{\text{rad}} \propto \left( \frac{dI}{dt} \right)^2
\end{equation}

\begin{figure}[!t]
\centerline{\includegraphics[scale=0.3]{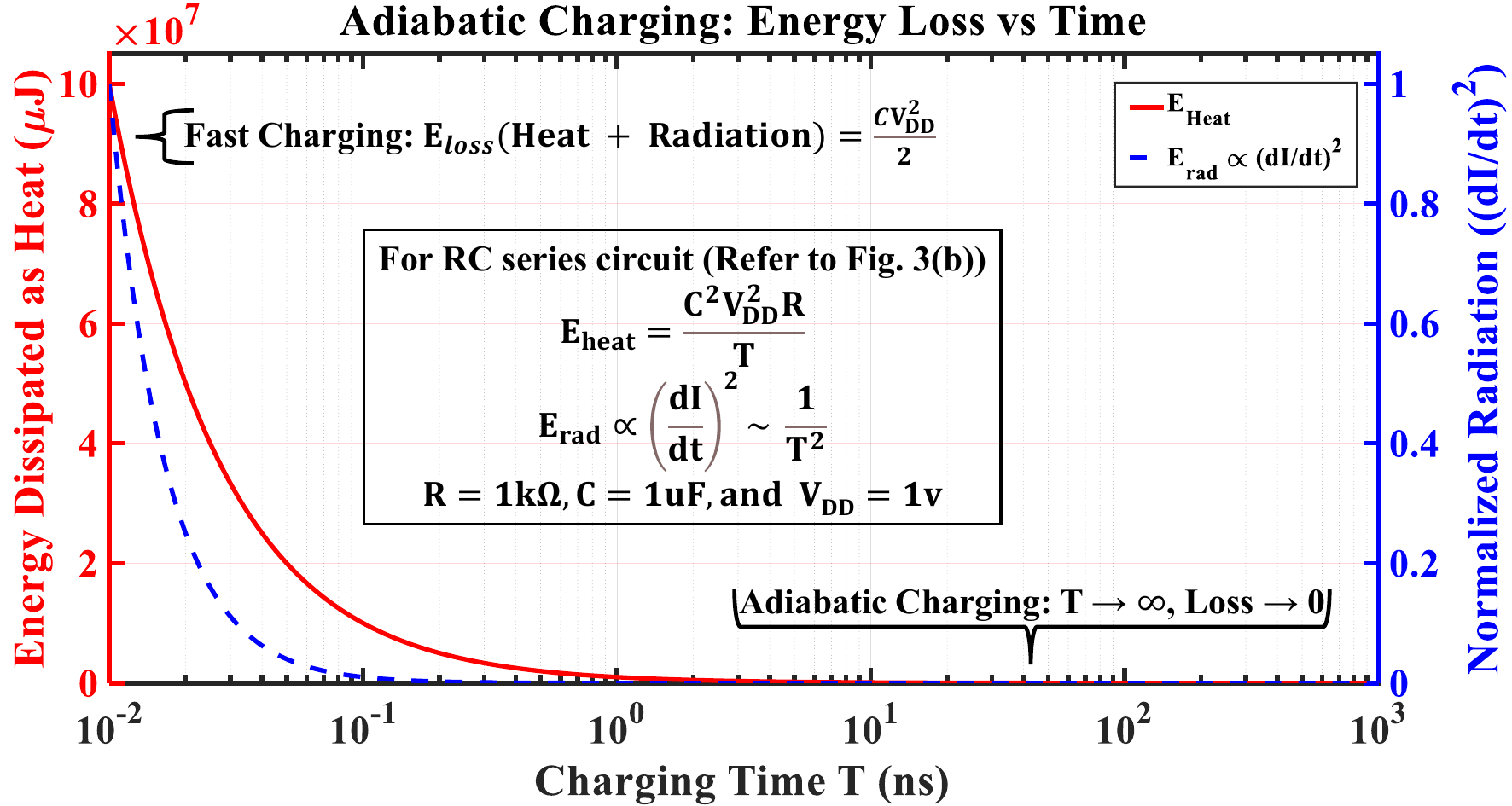}}
\vspace{-1mm}
\caption{Energy lost as heat and normalized EM radiation vs. charging time (log scale). As charging slows, energy losses approach zero, confirming that adiabatic charging suppresses EM side-channel leakage by reducing voltage transition rates.}
\vspace{-4mm}
\label{adiabatic}
\end{figure}

In practical scenarios, the time interval \( dt \) is much smaller than the current variation \( dI \), making \( \frac{dI}{dt} \approx \frac{1}{T} \). Consequently, the radiated energy is approximated as \( E_{\text{rad}} \propto \left( \frac{dI}{dt} \right)^2 \propto \frac{1}{T^2} \), indicating an inverse-square dependence of radiation on charging time. As \( T \to \infty \), the total energy loss \( (E_{\text{heat}} + E_{\text{rad}}) \to 0 \), resulting in minimal EM leakage. While resistance cannot always be minimized, unless superconducting materials are used, which suppress \( E_{\text{heat}} \) but not necessarily \( E_{\text{rad}} \), increasing the charging duration \( T \) effectively reduces \( dI/dt \), thereby lowering time-varying magnetic fields and radiated energy.

This behavior aligns with the Poynting vector relation \( \vec{S} = \vec{E} \times \vec{H} \propto \frac{dV}{dt} \cdot \frac{dI}{dt} \), where slower voltage and current transitions lead to reduced radiative power flow \cite{balanis2016antenna}. Thus, reducing both \( dV/dt \) and \( dI/dt \) using adiabatic charging minimizes the power flux through the circuit boundary, mitigating EM leakage. This trend is illustrated in Fig. \ref{adiabatic}, where both resistive and radiative losses decrease sharply with increasing charging time, validating the theoretical predictions. Hence, it is established that adiabatic charging is effective in reducing the \(\frac{1}{2} C V_{DD}^2\) energy loss and EM SCA attacks. Further improvement can be achieved when combined with other layout-level mitigation techniques discussed in Section~\ref{section2}.

\section{Conclusion and Future Work}
This paper presented a detailed mathematical and physical analysis of energy dissipation during capacitor charging in CMOS circuits, emphasizing both resistive and EM radiation losses. It also described how the switching of transistors results in EM emissions that contribute to the EM SCA attacks, leading to the recovery of the secret key from a cryptographic IC. To address this, adiabatic charging was proposed and mathematically validated as a strategy to reduce EM radiation losses and thereby provide EM SCA resilience. 

Future work will focus on the hardware implementation of adiabatic charging methods in standard CMOS circuits, with a focus on balancing security, energy efficiency, and circuit complexity. Additionally, this framework can be integrated with the existing side-channel countermeasures, which were discussed in~\cite{das2022sca, das2020and, das2019stellar} and in Section~\ref{section2}.

\bibliographystyle{IEEEtran}
{
\bibliography{bibliography}
}

\end{document}